# Shape it Better than Skip it: Mapping the Territory of Quantum Computing and its Transformative Potential

**Imed Boughzala**
*Institut Mines-Telecom Business School*
*France*                                    imed.boughzala@imt-bs.eu

**Nesrine Ben Yahia**
*RIADI Laboratory, National School of Computer Science*
*University of Manouba, Tunisia*            nesrine.benyahia@ensi-uma.tn

**Narjès Bellamine Ben Saoud**
*RIADI Laboratory, National School of Computer Science*
*University of Manouba, Tunisia*        narjes.bellamine.bensaoud@ensi-uma.tn

**Wissem Eljaoued**
*RIADI Laboratory, National School of Computer Science*
*University of Manouba, Tunisia*            wissem.eljaoued@ensi-uma.tn

## Abstract

Quantum Computing (QC) is an emerging and fast-growing research field that combines computer science with quantum mechanics such as quantum superposition and quantum entanglement. In order to contribute to a clarification of this field, the objective of this paper is twofold. Firstly, it aims to map the territory in which most relevant QC researches, scientific communities and related domains are stated and its relationship with classical computing. Secondly, it aims to examine the future research agenda according to different perspectives. We will do so by conducting a systematic literature review (SLR) based on the most important databases from 2010 to 2022. Our findings demonstrate that there is still room for understanding QC and how it transforms business, society and learning.

## 1. Introduction

In recent years, the field of Quantum Computing (QC) has rapidly progressed and emerged as one of the highly topical fields of research. In fact, QC has the potential to offer significant computational capabilities over classical (conventional) computing by exploiting the principles of quantum mechanics such as superposition (i.e. every quantum state can be represented as a sum of two or more other distinct states) and entanglement (occurs when a group of particles are generated, interact, or share spatial proximity in a way such that the quantum state of each particle of the group cannot be described independently of the state of the others). On this path, this important computational advantage will help to solve many complex and computational problems in several fields of research. Furthermore, it has attracted attention from both academics and industries alike.

Today, many big companies such as Google, D-Wave Systems, IBM, Honeywell, Rigetti Computing and Microsoft compete to build the first large-scale commercial quantum computer and research advancements in this new field ultimately will lead to quantum supremacy [1]. Complex problems that have been considered unsolvable by classical computers might be solvable with the emergence of quantum computers. In addition, solving computational problems may require billions of years in a traditional computational setting, while in theory, and in a quantum computational setting, they can be solved within a few hours [2]. According to [3], the complexity of any computing task performed using a quantum computer, can be decreased by at least a square root compared to the complexity of this task performed on the classical computer, and some exponentially complex problems can be solved in polynomial time (i.e. NP-completeness).

On the other hand, while its capabilities and benefits are seen in making magnificent research advancements in the field of computing, QC is still seeking to be shaped [4]. In fact, the emergence of QC has led to the apparition of quantum algorithms (such as factorization of integers and discrete logarithms [5]), quantum models (such as quantum neural networks [6]), quantum technologies [2] (such as quantum simulations, communication, computation, calculators, sensors and radars), and quantum applications (such as quantum cryptography, cyber security, meteorology, military, biology, energy, aeronautics, aerospace, etc.). Despite the abundance of literature (often technical), the field is in fact still unclear and its impact not well defined.

In this context, this paper tries to enhance the understanding by studying the main QC building blocks and its relationship with classical computing. In fact, the key motivation behind this study is to conduct a comprehensive and rigorous study of the current literature on QC based on a systematic literature review (SLR). It encompasses the definition of a QC, the mapping of its territory and its relationships with classical computing. This review will also discuss open challenges, implications and future QC.

To do so, this paper is organized as follows. In Section 2, we will present a WH-questions based background of QC. In Section 3, we will show the SLR conducted in this study. In Section 4, we will demonstrate our findings. Section 5 will discuss these results. In Section 6, we will highlight various open challenges and promising future directions for QC. Finally, in Section 6, we conclude the paper.

## 2. Background

In order to give a better understanding of the QC background, we will be based on the following six WH-questions:

**What is QC?** QC harnesses quantum mechanical phenomena to perform calculations and to boost computational efficiency in solving complex problems. It aims to process information in a fundamentally different way than traditional computing. Classical computers operate on binary bits and information processed in the form of zeros or ones. However, quantum computers transmit information via quantum bits, qubits [7] and information processed in the form of zeros or ones or both simultaneously.

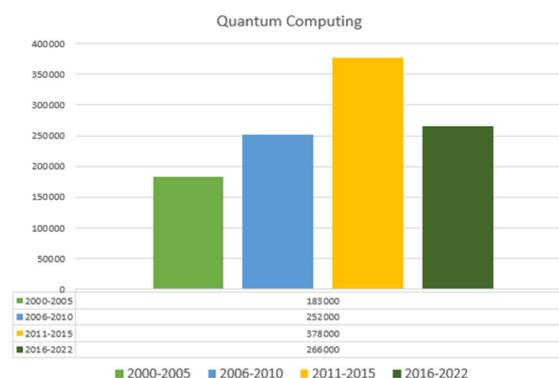

**Fig. 1**. Emergence of QC from 2000 to 2022

**When is QC emerged and raised up?** In 1982, [8] pointed out that the computer has to be working quantum mechanically to simulate a quantum system. In 1994, interest in QC rose dramatically when the first proposal for practical implementation of a QC was presented in [9] by developing a quantum algorithm, which could find the prime factors of large numbers efficiently. Since, QC has widely emerged and has become a fast-growing research field. In fact, from 2000 to 2022, there has been a significant increase in the number of scientific publications indexed in Google Scholar using this concept as shown in Figure 1.

**How do QC work?** Quantum superposition and quantum entanglement - "Qubits can be linked with other qubits in a process called entanglement." Superposition is what gives quantum computers speed and parallelism, meaning that these computers could theoretically work on millions of computations at once. When combined with superposition, quantum computers could process a massive number of possible outcomes at the same time.

**Why is QC used and emerged?** Actually, most experimental studies have relied on supercomputers which are very large classical computers often with thousands of classical CPU or GPU cores to solve complex problems. However, supercomputers don't have the working memory to hold the myriad combinations of real-world problems and they have to analyze each combination one after another, which can take a long time. In this context, the potential power of QC is based on the ability of enormous quantum parallelism to converge rapidly (quantum supremacy). In fact, in a QC setting, some exponentially complex problems can be solved in polynomial time which represents a remarkable increase in processing efficiency and time saved. Thus, QC can create vast multidimensional spaces in which to represent these very large and complex problems while classical supercomputers cannot do this.

**Where is QC used and applied?** According to Gartner1, QC does hold the potential to revolutionize many industries, including Cyber-security and Cryptography, Healthcare, Drug Design and development, Chemistry, Physics, Biology, Aeronautics/Aerospace, Finance and Cryptocurrency, Marketing, Transport, Energy and even Weather forecasting.

## 3.  Related Works

Despite QC research advancements, the map of the overall field is quite poor and we lack rigorous systematic reviews to better understand this field and the challenges accompanying it. However, some attempts have been made and there have been only a few surveys conducted on QC in the existing literature to address this deficiency. For example, [2] review the most recent results of QC technology and address the open problems of the field. They listed QC building blocks and the most recent research directions on the physical implementation of quantum devices, computers, and algorithms. Their findings indicated that QC technologies continue to hold tremendous potential for future computation, networking, and communications. The study of [4] is limited to review QC models, be it mathematical, algorithmic, circuit-wise, and machine-wise where their findings demonstrate that many of the existing models of QC are either mathematical or algorithmic.

In [10], authors outlined a survey and an analysis of QC basic concepts, some QC algorithms and prospects in the field of constructing a scalable quantum computer. Their results testified to the lack of sufficient progress in constructing a scalable QC device in terms of the implementation of well-known quantum algorithms. In [11], a survey on quantum programming languages is presented and it has been concluded that the area of compiling quantum programming languages has received relatively little attention. In [12], authors outlined a survey on quantum computers, quantum computer systems, and quantum simulators and they proposed to advance more intensive research in a physical realization of components of quantum computers and the architectural solutions for quantum computers as well as the (practical) quantum algorithms.

 As part of our related works, we found that most existing papers regarding QC survey focused on one angle from the scope (e.g. algorithms, programming, etc.), while this field actually covers a wider potential of scope. We believe that QC should also be studied from a large vision to better shape it by mapping its territory. Then, the relationship between quantum and classical computing should also be outlined and clarified and the impacts of QC on business, society and learning, etc. should be discussed. In addition, in order to rigorously explore these points, we will conduct a SLR that will be presented in the next section.

---

[1] https://itango.eu/the-cios-guide-to-quantum-computing/

## 4.    Methodology

In this paper, we intend to explore the shape of quantum computing by mapping its territory and to provide a future research agenda with new perspectives, based on a systematic literature review (SLR) method. Our study is based on the SLR process and guidelines proposed in [13] and this choice is motivated by its effectiveness to shape new emerging fields such as the shape of digital intelligence in [14].

Figure 2 summarizes the stages in our SLR into three main phases: Planning the Review, Conducting the Review, and Reporting the Review.

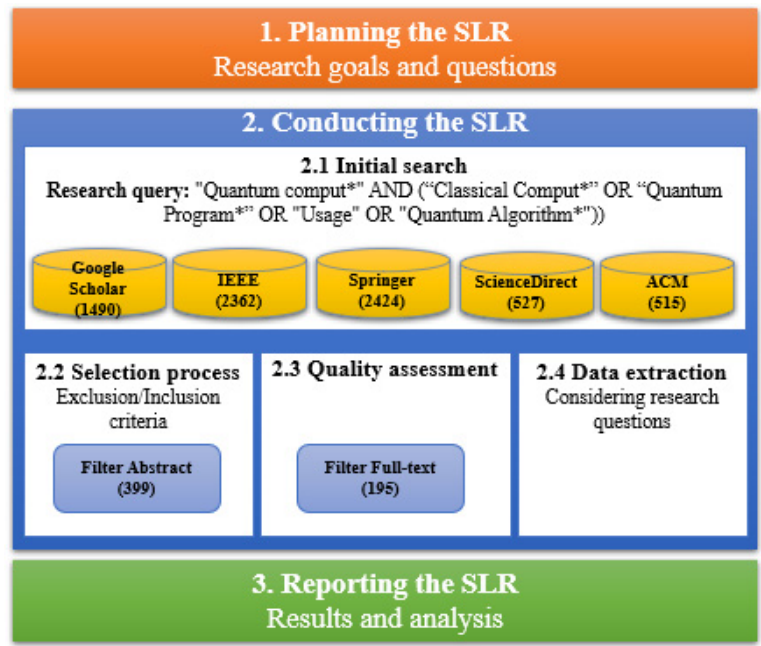

**Fig. 2.** Our SLR process

### 4.1.    Planning the Review

The main goal of this first step is to identify the need for a review by specifying the research goals and question(s). In our case, our research goals are to understand quantum computing and its relationship with classical computing and how it will transform business and society. Based on the mentioned goals, we raise the following research questions (RQs):

- **RQ1.** What are the definitions of QC in the existing literature?
- **RQ2.** What is the scope/territory of QC?
- **RQ3.** What is the nature of relationships between classical and quantum computing (match, transform, complete ...)?
- **RQ4.** How QC will transform business, society and learning?

### 4.2.    Conducting the Review

This second step aims to select primary studies, study quality assessment, extract and synthesize data.

**Search Strategy**

The systematic search of primary studies was conducted from 2010 to 2021 and was performed on five electronic databases that included IEEE Xplore, Google Scholar, ACM Digital Library, Springer, and ScienceDirect. These databases have been suggested by [13] as they constituted the search engines typically used in Computer Science / Software Engineering and Information Systems SLR. Then, we used the following logical research request:
"Quantum comput*" AND ("Classical Comput*" OR "Quantum Program*" OR "Usage" OR "Quantum Algorithm*"))

This search request was successfully executed on all databases and the searches returned a total number of 7318 results from the five databases consulted as seen in Table 1. Then, we filtered results by keeping only journal papers and removing duplicated papers which generated a total of 1735 papers.

**Table 1.** Selected papers per journal

| Database | Papers number | Journal papers |
|---|---|---|
| Google scholar | 1490 | 143 |
| IEEE | 2362 | 635 |
| Springer | 2424 | 401 |
| science direct | 572 | 447 |
| ACM | 515 | 111 |
| **Total** | **7318** | **1737** |
| **Total (no doubles)** | | **1735** |

Thus, according to the growth in the number of results retrieved, we recognize that QC is an emerging field. However, given the exploratory nature of much of the research in this field, in order to identify and select the relevant studies and to provide more consistency and meaningfulness, we decide to consider in this primary search phase the "basket of eight journals" that represents the top journals in the Information Systems (IS) field[2]. Our search request was executed on these eight journals and the searches returned only two results as illustrated in Table 2.

**Table 2.** Papers number for the basket of 8 IS journal

| Basket of 8 IS Journals | Papers number |
|---|---|
| European Journal of Information Systems | 0 |
| Information Systems Journal | 0 |
| Information Systems Research | 0 |
| Journal of AIS | 0 |
| Journal of Information Technology | 2 |
| Journal of MIS | 0 |
| Journal of Strategic Information Systems | 0 |
| MIS Quarterly | 0 |
| **Total** | **2** |

Then, in order to find additional primary studies, we reflect also on the selection of the top eight journals in computer science by searching the first eight journals ranked A* in Computing Research and Education (CORE) Portal Journal[3]. Thus, we selected the top eight A*ones by excluding the eight journals of IS basket. Then, our search request was executed again on these eight CS journals and the returned results are shown in Table 3.

**Table 3.** Papers number of top A* eight CS Journals

| Top A* eight CS Journals | Papers number |
|---|---|
| ACM Computing Surveys | 16 |
| ACM Transactions on Computer Systems | 0 |
| Algorithmica | 13 |
| Annual Review of Information Science and Technology | 0 |
| IEEE Transactions on Computers | 21 |
| IEEE Transactions on Information Theory | 109 |
| Information Systems | 0 |
| Journal of Computer and System Sciences | 7 |
| **Total** | **166** |

Furthermore, resorting to the same analogy as the basket of eight IS journals, we also decide to select the first eight journals that have more published papers related to our request string from the five consulted digital databases (Figure 2). It is worth noting here that the eight journals of IS basket and the top eight A* CS journals have been excluded to avoid papers duplication.

**Table 4.** Papers number of the top 8 publishing Journals

| Top 8 publishing Journals | Papers number |
|---|---|
| IEEE Access | 93 |
| Science china information sciences | 40 |
| IEEE Transactions on Image Processing | 34 |
| IEEE Transactions on Quantum Engineering | 39 |
| IEEE Transactions on Neural Networks and Learning systems | 25 |
| Theoretical Computer Science | 31 |
| Natural Computing | 23 |
| IEEE Transactions on Cybernatics | 21 |
| **Total** | **306** |

### Selection process

In this step, we the defined selection criteria to select relevant studies and to determine which studies are included or excluded. Studies that met the following criteria were included.

**Table 5.** Selection criteria

| Exclusion criteria | Inclusion criteria |
|---|---|
| C1. Duplicated papers | C5. Published between 2010 and 2022 |
| C2. Conferences, Books, posters | C6. Limited to the fields Computer science or Information systems |
| C3. Not written in English | C6. Focus on QC scope |

In this step, irrelevant studies were removed based on the inclusion/exclusion criteria. Finally, a total number of 399 papers passed these criteria.

### Quality assessment

Next, in order to ensure the accuracy and reliability of the final study selection, a careful quality assessment step was performed. First, Zotero, a reference manager, was used to share the selected papers between authors. Second, the 399 selected papers were distributed among the four authors. Third, each co-author accurately read each paper independently and tried to assess its reliability and then made the decision on whether or not to include it as a relevant paper based on the quality assessment questions presented below.

In fact, the quality assessment questions are defined to evaluate the rigor and credibility of the selected articles. The evaluation requires the complete review of the paper. Based on the works of [15], [16], and [17], we defined the following quality criteria stated as questions:

- − Is there an adequate description of the context in which the research was carried out?
- − Are there a clear statement of research aims? Does the paper describe an explicit research question?
- − Is the research design appropriate to address the research aims?
- − Is the literature review adequate? Is the collected data addressing the research issue?
- − Is the data analysis sufficiently rigorous?
- − Is there a clear statement of findings?
- − Is the study valuable for research or practice?
- − Does the paper discuss limitations or validity?

Each question has four possible options: (0) issue is not mentioned at all, (1) little mentioned, (2) adequately addressed and (3) completely addressed [15]. Hence, we used a four points Likert scale for collecting answers. Articles with an average quality score lower than 1, were removed. At the end of this process step 195 papers from the 399 were qualified to be analyzed for the data extraction step. The list of these 195 papers is accessible via this link.

**Data Extraction**

In this step, we extracted data from the qualified articles. In order to converge rapidly and to provide a structured approach for the review, each author worked independently to extract data from all primary studies, guided by an extraction form. We designed this data extraction form by considering our research questions.

## 5. Results and Analysis

The goal of this paper is to systematically review and synthesize the state of the art in the related area and to get an explicit view of QC research including the recent trends and directions in this field to identify research gaps for future study.

As a first result, we present in Figure 3 the distribution of the 195 papers by journal.

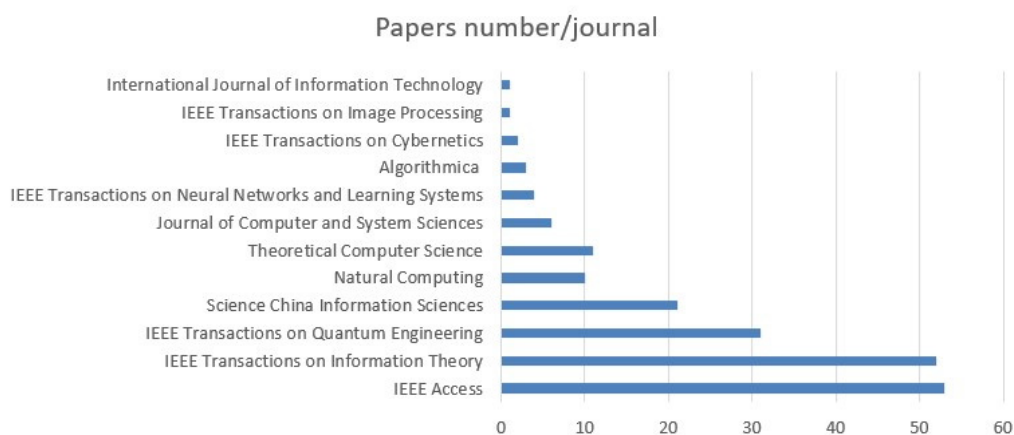

**Fig. 3.** Distribution of selected papers by journal.

Then, we present in Figure 4 the distribution of papers per year where we see the expansion in QC research especially in three last years.

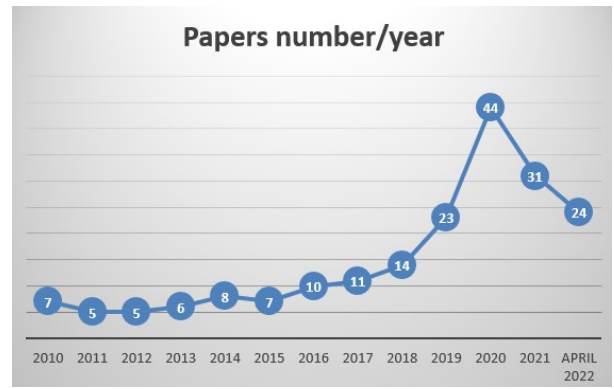

**Fig. 4.** Distribution of selected papers per year.

Next, in order to map the territory of QC and to understand its scope, Figure 5 illustrates the distribution of the 195 papers per treated scope. We note here that the different areas that have been treated when dealing with QC are as follows: quantum algorithmic, quantum programming, architecture (hardware, processors, circuits, etc.) and quantum networks (quantum communication, protocols, routers, etc.).

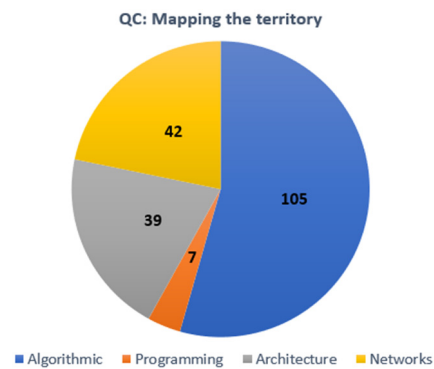

**Fig. 5.** Mapping the territory of QC

As illustrated in Figure 5, quantum algorithm presents the most advanced area for quantum computing compared to the others areas (105 out of 195 i.e. 53,58%). Thus, we decide to zoom in on this area to understand its sub-areas. We note that the Quantum algorithm area can also be categorized into five areas that are Quantum Computation (such as quantum query complexity studied in papers number 3, 65, 119 and 150), Quantum Optimization (Evolutionary algorithms and quantum-inspired genetic algorithm and quantum-inspired metaheuristics studied in such as papers 1, 14, 33 and 42), Quantum security (encryption and post-quantum cryptographic algorithms presented in 11, 19, 32, and 96), Quantum machine learning (quantum neural networks studied in 4, 44, 91 and 101) and Quantum image processing (studied in 30,

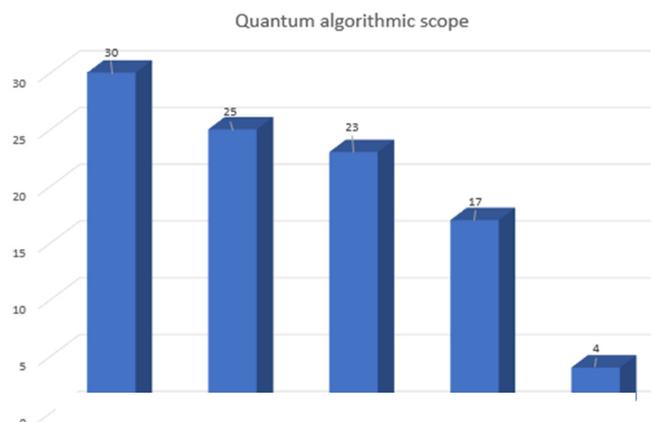

**Fig. 6.** Quantum algorithm sub-areas

116, 128 and 139).

With regard to our RQ3, 28 papers from the 195 selected ones deal with the relationship between quantum computing and classical computing. Papers 1-5 in this list all share the same insight by considering QC as an accelerator that can speed up classical computing. In this context, paper 6 confirms that quantum computing has a promising extension to classical computing and it can provide more computational power in solving some complex problems. Furthermore, paper 30 announces that QC can achieve a significant exponential acceleration. Paper 14 suggests the hybridization of quantum-inspired algorithms with classical algorithms in a quantum-inspired genetic algorithm for resource-constrained project-scheduling and paper 35 deals with the adaptation of quantum and traditional networks into a hybrid paradigm. Moreover, other papers, such as 8 and 33, state that quantum intelligent optimization algorithms are new kind of algorithms where quantum versions are at least as good as their classical version counterpart. Paper 58 contemplates that on the one hand quantum communications can replace classical communications but on the other hand classical communication can "replace" quantum communication when the advantage of using quantum communication can be at most (quasi-) polynomial in terms of complexity.

Finally, with regard to our RQ that concerns the impact of QC on business, society and learning and their potential transformation, we note a knowledge gap which is not bridged yet.

## 6. Discussion

Through this literature review, we can assume that most of the research effort yet is related to the design science (technical aspects) rather than behavioral science (usage aspects) [18] . If RQ2 and RQ3 are already covered in this review, we can notice RQ1 and RQ4 are not.

### 6.1. Definition of QC

From the 195 selected papers, no one gives a clear definition of QC. According to Wikipedia, "QC is the exploitation of collective properties of quantum states, such as superposition and entanglement, to perform computation. The devices that perform quantum computations are known as quantum computers." Exploring other references, we decide to build this table to see if there are any complementary elements and provide their own definition:

**Table 6.** Definitions of QC

| Reference | Definition |
|---|---|
| Investopedia.com | is an area of computing focused on developing computer technology based on the principles of quantum theory (which explains the behavior of energy and material on the atomic and subatomic levels). Quantum computing… uses quantum bits or qubits. It harnesses the unique ability of subatomic particles that allows them to exist in more than one state (i.e., a 1 and a 0 at the same time). |
| IBM.com | harnesses the phenomena of quantum mechanics to deliver a huge leap forward in computation to solve certain problems. |
| Gartner.com | is a type of non-classical computing that operates on the quantum state of subatomic particles. The particles represent information as elements denoted as quantum bits (qubits). A qubit can represent all possible values simultaneously (superposition) until read. Qubits can be linked with other qubits, a property known as entanglement. Quantum algorithms manipulate linked qubits in their undetermined, entangled state, a process that can address problems with vast combinatorial complexity |

To respond to RQ1, according to us, QC is the study of how to use the principles of quantum theory to create new ways of computing, software, hardware and network, all at a much higher performance than their classical counterparts for specific use cases.

### 6.2. The Transformative Potential of QC

To fill in the gap of RQ4 that concerns the impact of QC on business, society and learning and their potential transformation, we decided to refer to what is digital transformation and how QC could be impactful.

Digital transformation is "a radical or incremental change process that starts with the adoption and use of digital technologies, then evolves into an implicit holistic transformation of an organization, or deliberate to pursue value creation." [14]. It operates on three layers: User

eXperience (UX), Operational Processes (OP) and Business Models (BM). Digital transformation, also known as **digitalization**, refers to a business model driven (approach) by "the changes associated with the application of digital technology in all aspects of human society" [19]. It is usually implemented through **digitization**, i.e. the "ability to turn existing products or services into digital variants, and thus offer advantages over tangible product" [20]. QC as an emerging technology could digitally transform business and society on the three layers and could have different implications related to different transitions.

### 6.2.1 UX

"The user experience (UX or UE) is how a user interacts with and experiences a product, system or service. It includes a person's perceptions of utility, ease of use, and efficiency. Improving user experience is important to most companies, designers, and creators when creating and refining products because negative user experience can diminish the use of the product and, therefore, any desired positive impacts; conversely, designing toward profitability often conflicts with ethical user experience objectives and even causes harm." (Wikipedia)

QC would enhance the UX of stakeholders by giving many advantages, among others, to the User, Developer and Manager. To the user, it offers to compute more speed navigating on internet, for example, to the developer, it offers more alternatives to solve complex problems and to the manager (decision maker), it enables the ability to understand/anticipate harmful situations (such as cyberattacks).

### 6.2.2 OP

A business or operational process is an organized set of activities or tasks that produces a specific service or product. It refers to the essential business activities that deliver value to the customer.

QC holds the potential to revolutionize many industries and business activities by enhancing the effectiveness and efficiency of the specification, design, execution and delivery business processes.

### 6.2.3 BM

A business model describes the rationale of how an organization creates, delivers, and captures value, in economic, social, cultural or other contexts. The process of business model construction and modification is also called business model innovation and forms a part of business strategy." (Wikipedia)

QC would give more openness to transform the current business models. The power of data and data-driven business models will take more importance. QC will give many possibilities to extend the power of data science and artificial intelligence (i.e. Quantum Lachine/Deep Learning).

## 7. Future Research Agenda

To define a future research agenda that could be useful to prioritize our research actions and to give perspectives to others, we have built this table by addressing on the one hand the dimension and on the other the potential implication and questions.

**Table 7**. Future research agenda

| Dimension | Implication/questions |
|---|---|
| Social | QC (devices, computers, algorithms and others) will call to new competencies (hardware, software and network). What are the skills to acquire? What kind of new jobs to be created? This also could aggravate the digital divide by the emergence of new forms. Will QC participate in the augmented intelligence or diminished intelligence of citizens? |
| Economic | QC will open up new markets and develop new businesses. Will the classical computing sector be radically transformed? Will the competitiveness and leadership of businesses or countries be affected? Several sectoral applications will give rise to several business opportunities. What are the specific use cases? |

| Technological/technical | QC holds the potential to revolutionize the computing and telecommunication sectors: interfaces, software, hardware and network. Will the classical computing scope be called into question? Or certain dimensions only? What are the most technical challenges? |
|---|---|
| Ecological/energetic | QC would reduce the carbon footprint thanks to the quantum computer. These computers are supposed to consume very little electricity thanks to the quantum characteristics of physical matter. To what extent digital sobriety is encouraged by QC? Is there a way to measure / assess that? |
| Education and learning | QC would make use of several adaptations in the educational approaches and curricula. To what extent computer literacy would be revolutionized?<br>However, several technical and managerial problems will appear: interface between classical computing and QC, adoption of QC by organizations, rationalization of the choice of applications, etc. |
| (geo)Political | Digital sovereignty is for states to keep a technological & scientific advance in order to remain among the leaders of the world while being master of their own destiny by promoting cutting-edge research and technological innovation (Patent filing, Nobel Prize, etc.). This relates to digital intelligence, digital infrastructures, data sphere, safety of people and places, etc. QC like AI, Cloud/Edge computing, IOT, 6G... is the new area in which developed countries must keep their sovereignty to avoid dependence on others. Several billion dollars are spent by the US and China in this technological race. Are there any QC dimensions more important than others in this race? |

## 8. Threats of Validity

At this stage, we assume that there might be some validity threats with our research findings, and we try here to self-assess it in order to denote the trustworthiness of our results, to what extent they are true and not biased from our subjective point of view? Furthermore, these potential threats will be addressed according to the classification proposed in [21].

Regarding the construct validity, we assume that the research coverage of primary studies could be limited to a particular field. However, we have used in this research, five electronic databases (IEEE Xplore, Google Scholar, ACM Digital Library, Springer, and ScienceDirect) that have been suggested by [13] as they constituted the search engines typically used in Computer Science and Software Engineering SLR. We also considered both computer science and information systems fields to study QC from different angles to collect more insights on this emerging field. The maturity of the QC field could be another factor that can affect construct validity, however, we believe that more than 10 years (from 2010 to May 2021), is enough time to review this emerging research field in a systematic way. Regarding the internal validity, there might be some issues regarding the selection and quality assessment of the primary studies that could be biased regarding the authors' expected results. However, to minimize this threat, our quality assessment criteria have been objectively defined based on the works of [15], [16], and [17]. Finally, concerning the external validity, that concerns the degree to what extent the results can be generalized to other search contexts, we assume that there might be some issues regarding generalization of our SLR process. To reduce this threat, we adopted a search process from the well-established guidelines of [13] that is based on three phases (planning, conducting and reporting the review) and that has been widely used in SLR works.

## 9. Conclusion

Motivated by the growing significance of QC, we decided to systematically gather and rigorously analyze and synthesize the literature, in order to explore how the literature is. Our primary goal was to map the territory of QC. It also intends to analyze and discuss the potential transformation of QC for business, society and learning. This SLR promises several potential benefits for both researchers and practitioners. The findings and discussion are potential value for researchers to shape their future research directions according to different dimensions and even disciplines. This research indicates that despite the imperative role of QC as emerging technology for digital transformation and innovation, it is not comprehensively investigated by academic literature from the behavioral – usage - viewpoint. However, practical research, innovation and investment, such as those done by companies like IBM, Microsoft and others, more widely countries, show the important of QC through white papers, websites and professional magazines.